\documentclass[referee]{raa}           

\usepackage{graphicx,times}
\usepackage{natbib}
\usepackage{amssymb,amsmath}
\bibpunct{(}{)}{;}{a}{}{,}

\usepackage[pagebackref=true]{hyperref}

\begin{document}

   \title{Precise Interstellar Extinction Estimates for Nearly Ten Million Stars Based on Multiple Spectroscopic Surveys}

 \volnopage{ {\bf 20XX} Vol.\ {\bf X} No. {\bf XX}, 000--000}
   \setcounter{page}{1}

   \author{Bao-Kun Sun
   \inst{1}, Qiu-Ling Yan\inst{1}, Qing-Guo\inst{1}, Jin-Hua Yang\inst{1}, Yan-Hui Chen\inst{1}, Zheng-Yu\inst{1}, Bai-Song Zhang\inst{2}
   }
   \institute{$^1$Institute of Astrophysics, Chuxiong Normal College, Chuxiong 675000, China\\ 
   $^2$School of Integrated Circuits, Dezhou Vocational and Technical College, Dezhou, Shandong 253034, China\\ {\it sunbaokun@cxtc.edu.cn; yuzheng\_hs@163.com; zhangbaisong@mail.ynu.edu.cn}\\
\vs \no
   {\small Received 20XX Month Day; accepted 20XX Month Day}
}

\abstract{
Accurate interstellar extinction estimates are vital for revealing the intrinsic physical properties of stars and the dynamic structure of the Milky Way. Based on multiple optical and near-infrared spectroscopic surveys---including LAMOST, APOGEE, GALAH, Gaia-ESO, SEGUE, and DESI---we use MLPs and polynomial regressions to calibrate giants and dwarfs separately onto the APOGEE parameter scale, constructing a combined spectroscopic catalog of nearly ten million stars with calibrated stellar atmospheric parameters. This catalog not only provides a robust dataset for our extinction measurements, but also serves as a reference sample for estimating stellar parameters from photometric surveys. Both cross-survey comparisons and external comparisons with high-resolution spectroscopic benchmarks show good agreement in the calibrated parameters. By applying the star-pair method to this extensive sample, we obtain precise extinction estimates. Incorporating distance information, we characterize the spatial distribution of extinction covering over 33,000 square degrees of the sky. Comparisons with existing literature demonstrate that our extinction estimates achieve an exceptional precision of 0.019--0.022~mag. The value-added catalog and open-source Python pipeline for line-of-sight extinction released in this work will provide robust support for future Galactic structure studies and photometric calibrations.
}

   \authorrunning{B.-K. Sun et al. }            %author_head in even pages
   \titlerunning{Precise Extinction for 10M Stars}  % title_head in odd pages
   \maketitle

%________________________________________________ sections below
% 
\section{Introduction}           %% first-level sections will be auto-capitalized
\label{sect:intro}
Interstellar dust grains are ubiquitously distributed across diverse astronomical environments  \citep{chen2024three}. Although they constitute only about 1\% of the mass of the interstellar medium (ISM), or roughly 0.1\% of the total mass of the Milky Way, they absorb approximately 30\% of the total electromagnetic radiation energy in the Galaxy \citep{draine2003interstellar}. The resulting extinction and reddening effects, driven by the absorption and scattering of light by dust grains from ultraviolet to near-infrared wavelengths, significantly alter the observed brightness and colors of celestial objects. This poses a serious obstacle to the accurate assessment of their intrinsic physical parameters, such as effective temperature and metallicity \citep[e.g.,][]{sun2025photometric, huang2023beyond, zhang2023parameters, huang2025stellar}. Particularly in low Galactic latitude regions—where stellar density is high and dust obscuration is severe—interstellar extinction presents a critical bottleneck for studying the three-dimensional structure, formation, and evolution of the Milky Way if precise corrections are not applied \citep[e.g.,][]{chen2013three, green2015three, chen2017constraining}.

Traditional two-dimensional (2D) dust maps are widely constructed either by modeling the far-infrared thermal emission to trace column density \citep[e.g.,][]{schlegel1998maps, abergel2014planck} or by measuring stellar near-infrared color excesses \citep[e.g.,][]{lombardi2001smooth, lombardi2009nicest}. While recent refinements have successfully suppressed systematic errors such as cosmic infrared background contamination \citep{chiang2023corrected} or improved angular resolution using neutral hydrogen \citep{lenz2017new}, these maps inherently present only the total integrated extinction projected along a given line of sight. Consequently, they lack distance information and fail to resolve the line-of-sight depth of distinct dust structures. This projection effect poses a critical challenge for studies within the Milky Way disk, where stars embedded inside dust layers rather than situated behind them suffer from a severe overestimation of their actual extinction if a 2D total correction is applied.

To overcome these severe projection effects, introducing distance information to transition from 2D projections to three-dimensional (3D) extinction mapping is critical for studies within the Galaxy. Early pioneering works often relied on Galactic stellar distribution models to break the distance-extinction degeneracy, as exemplified by \citet{marshall2006modelling}, who mapped the inner Galactic plane using the Two Micron All Sky Survey (2MASS; \citealt{skrutskie2006two}) photometry combined with the Besançon model. Fueled by recent large-scale multi-band surveys and precise Gaia astrometry \citep{brown2018gaia}, modern 3D dust mapping has largely shifted toward data-driven and machine learning techniques to cover extensive sky areas. For instance, \citet{green20193d} integrated Pan-STARRS 1 \citep{chambers2016pan} and 2MASS data with Gaia parallaxes to map the northern sky, while \citet{guo2021three} utilized the SkyMapper Southern Survey (SMSS; \citealt{wolf2018skymapper}), 2MASS, The Wide-field Infrared Survey Explorer (WISE; \citealt{wright2010wide}) and Gaia data to construct a complementary 3D extinction map of the southern sky. Concurrently, \citet{chen2019three} developed a machine learning framework trained on spectroscopic surveys to characterize the Galactic disk. Most recently, the extensive low-resolution XP spectra from Gaia DR3 \citep{vallenari2023gaia} have ushered in a new era of all-sky 3D mapping with unprecedented statistics, enabling high-resolution parameter and extinction extractions via data-driven forward modeling \citep[e.g.,][]{zhang2023parameters, zhang2025three} or joint stellar parameter inference frameworks \citep[e.g.,][]{an2024blueprint}.

To extract highly precise extinction from these expanding spectroscopic datasets, the "star-pair" method has emerged as a powerful technique for 3D Galactic dust mapping \citep{yuan2013empirical, wang2025all}. By assuming that stars with near-identical atmospheric parameters possess indistinguishable intrinsic colors, this method derives extinction directly from observed color differences, fundamentally bypassing the systematic bottlenecks of traditional single-star SED fitting. Currently, the Large Sky Area Multi-Object Fiber Spectroscopy Telescope (LAMOST; \citealt{zhao2012lamost, liu2013lss}) provides a premier robust dataset for this approach by accumulating millions of high-precision stellar parameters. Furthermore, the synchronized influx of other massive spectroscopic catalogs—including the Apache Point Observatory Galactic Evolution Experiment (APOGEE; \citealt{majewski2017apache}), the Galactic Archaeology with HERMES project (GALAH; \citealt{de2015galah}), Sloan Extension for Galactic Understanding and Exploration (SEGUE; \citealt{yanny2009segue}), and the Dark Energy Spectroscopic Instrument (DESI; \citealt{desi2026data}) has significantly improved both sample volumes and parameter accuracies, expanding the star-pair method from local applications into a dominant tool for large-scale, high-resolution 3D Galactic mapping.

The structure of this paper is organized as follows. Section 2 introduces the spectroscopic datasets employed in this work and describe the methodology used to correct the systematic errors among different surveys. Section 3 delineates how the star-pair method is applied to this cross-calibrated spectroscopic sample to derive their precise color excess values. Section 4 details the spatial distribution of interstellar extinction, outline the procedure for calculating the $E(B-V)$ values based on this sample, and present a comprehensive comparison between our $E(B-V)$ estimates and those from previous studies. Finally, Section 5 offers a summary of our primary findings and concluding remarks.

\section{Data and Methodology} \label{data and Methodology}

In this study, we utilize six spectroscopic surveys, including LAMOST DR13, APOGEE-2, GALAH DR4, Gaia-ESO DR5, SEGUE DR12, and DESI DR1. To establish a consistent stellar parameter scale, multilayer perceptrons (MLPs) \citep{pinkus1999approximation} and polynomial regressions are applied to correct the systematic differences in stellar atmospheric parameters across these individual surveys separately for giants and dwarfs, calibrating them all to the reference scale of APOGEE. After removing duplicate sources, we ultimately construct a comprehensive value-added catalog containing approximately 10 million unique stars.

\subsection{Spectroscopic Data} \label{subsec:spectroscopic Data}

Our primary spectroscopic dataset is drawn from the LAMOST DR13 release, which comprises nearly 30 million spectra. As a unique quasi-meridian reflecting Schmidt instrument, the LAMOST telescope is capable of simultaneously obtaining spectra for up to 4000 celestial objects within a $5\degr$ diameter field of view, delivering a spectral coverage from 3,700 to 9,100\,$\AA$ with a resolution of $R \approx 1800$ in the low-resolution mode. The medium-resolution mode covers 4,950--5,350\,$\AA$ and 6,300--6,800\,$\AA$ in the blue and red arms, respectively, with a resolution of $R \approx 7500$ \citep{liu2020lamost}. For our star-pair algorithm, we specifically utilize the stellar atmospheric parameters—including effective temperature ($T_{\rm eff}$), surface gravity ($\log g$), and metallicity ($\rm [Fe/H]$)—from the Low-Resolution Spectroscopic Survey (LRS) and the Medium-Resolution Spectroscopic Survey (MRS) Stellar Parameter Catalogues of A, F, G, and K Stars. These two catalogs provide high-quality atmospheric parameters for 8,981,588 and 2,135,316 stars, respectively, which are derived by the official LAMOST Stellar Parameter Pipeline (LASP) with typical errors of approximately 150\,K, 0.25\,dex, and 0.15\,dex, respectively \citep{wu2011automatic, luo2015first}.

To significantly expand the overall sample size, sky coverage, and stellar parameter space, we additionally incorporate spectroscopic data from five other major international surveys: APOGEE-2, GALAH DR4, Gaia-ESO DR5, SEGUE DR12, and DESI DR1.
APOGEE-2 obtains high-resolution ($R \approx 22{,}500$) $H$-band spectra and provides stellar parameters through the APOGEE Stellar Parameter and Chemical Abundances Pipeline (ASPCAP; \citealp{garcia2016aspcap}), whose parameter scale serves as the reference for our cross-survey calibration.
In GALAH DR4, neural-network interpolation of synthetic spectra is used to derive stellar parameters from high-resolution ($R \approx 28{,}000$) HERMES optical spectra \citep{buder2025galah}.
The homogenized parameters in Gaia-ESO DR5 are based on VLT/FLAMES GIRAFFE ($R \sim 20{,}000$) and UVES ($R \sim 47{,}000$) spectra \citep{worley2024gaia}.
For SEGUE DR12, atmospheric parameters are determined by the SEGUE Stellar Parameter Pipeline (SSPP; \citealp{lee2008segue}) from optical spectra covering 3,800--9,200\,$\AA$ at $R \sim 2000$.
We also adopt stellar parameters from the DESI DR1 Milky Way Survey value-added catalog \citep{desi2026data}, derived from spectra covering 3,600--9,800\,$\AA$ at $R \sim 2000$--$5500$.
Together with LAMOST, this compilation essentially encompasses the vast majority of mainstream optical and near-infrared spectroscopic surveys currently available to the astronomical community. Crucially, because our methodology primarily leverages the relative ranking of stars within the $T_{\rm eff}$--$\log g$--$\rm [Fe/H]$ parameter space to select perfectly matched pairs, the final color excess determinations are highly resilient against potential systematic uncertainties in these parameters across different surveys \citep{chen2024photometric, yuan2013empirical} The sample selection criteria and the numbers of stars retained from the individual surveys are presented in Section~\ref{subsec:correction of systematic differences}.

\subsection{Correction of Systematic Differences} \label{subsec:correction of systematic differences}

However, combining multiple spectroscopic datasets introduces severe challenges due to systematic differences in stellar parameter estimations, which inherently arise from variations in instrumentation, spectral resolution, and data reduction pipelines across the surveys \citep{turchi2025survey}. To reduce differences between the parameter scales, we use sources in common with APOGEE to calibrate the other five surveys, treating LAMOST LRS and LAMOST MRS as separate datasets. To establish a robust and clean training sample for this cross-calibration, a stringent quality-cut framework is applied to each dataset. For all six surveys, we uniformly require baseline parameter uncertainties of $\sigma_{T_{\rm eff}} < 100$\,K, $\sigma_{\log g} < 0.15$\,dex, and $\sigma_{\rm [Fe/H]} < 0.1$\,dex. Since APOGEE serves as our reference scale, stricter constraints are imposed on its catalog, filtering sources with $\sigma_{T_{\rm eff}} < 50$\,K, $\sigma_{\log g} < 0.1$\,dex, $\sigma_{\rm [Fe/H]} < 0.05$\,dex, $\mathrm{SNR} > 100$, and keeping only stars satisfying the adopted quality-flag criterion with $\text{ASPCAPFLAG} == 0$ \citep{garcia2016aspcap}. For the remaining surveys, the specific quality cuts are applied through the following criteria:
\begin{enumerate}
\item LAMOST: Requires a $g$-band signal-to-noise ratio snrg $>$ 30 and a metallicity threshold of $\rm [Fe/H] > -2.5$\,dex.
\item GALAH: Requires reliable parameter flags with $\text{flag\_fe\_h} == 0$ and $\text{flag\_sp} == 0$.
\item Gaia-ESO: Excludes degraded sources where the quality flags equal $\text{SFLAG} == \text{SNR}$, $\text{SRP}$, or $\text{PSC}$.
\item DESI: Restricts the sample to confirmed stars by setting $\text{RR\_SPECTYPE} == \text{'STAR'}$.
\item SEGUE: Employs the recommended values for atmospheric parameters ($T_{\rm eff}$, $\log g$, and $\rm [Fe/H]$) and restricts the sample to higher quality spectra with $\text{ACF2\_SNR} > 20$.
\end{enumerate}
Only stars simultaneously satisfying these criteria are selected to construct the cross-calibration sample. After applying these stringent selection cuts, the final numbers of common sources shared between APOGEE and the other individual datasets—namely LAMOST LRS, LAMOST MRS, GALAH, ESO, SEGUE, and DESI—are 44,950, 25,818, 11,758, 568, 316, and 2,814, respectively.

\begin{figure}[!htbp]
\centering
\includegraphics[width=0.68\textwidth]{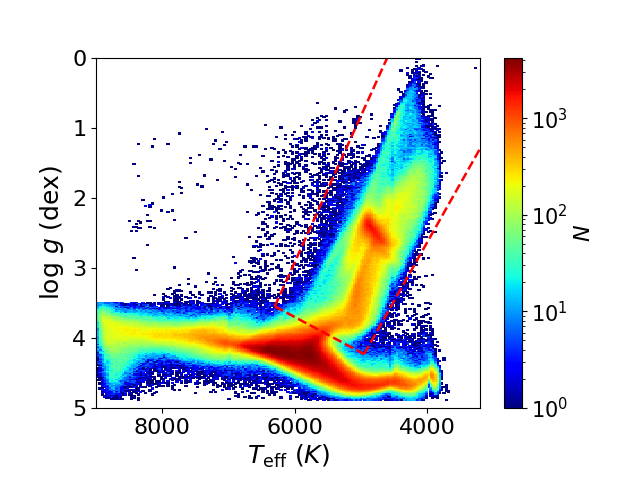}
\caption{Empirical separation of giants and dwarfs in the LAMOST LRS $T_{\rm eff}$--$\log g$ diagram. The color scale gives the number of stars per two-dimensional bin, $N$. The red dashed lines enclose the giant region in the upper right, and the remaining stars are classified as dwarfs.}
\label{fig:giant_dwarf_boundary}
\end{figure}

To account for population-dependent systematic differences, we first separate giants and dwarfs in the $T_{\rm eff}$--$\log g$ diagram \citep{sun2025photometric, guo2021three, niu2023internal}. We define an empirical boundary from the LAMOST LRS distribution in Figure~\ref{fig:giant_dwarf_boundary}. Stars in the upper-right region enclosed by the red dashed lines are classified as giants, and the remaining stars as dwarfs. We then fit separate calibration models for the two groups in each dataset. For LAMOST LRS, LAMOST MRS, GALAH, and DESI, we use MLPs. For each survey and stellar group, three MLP regressors are trained to predict the APOGEE values of $T_{\rm eff}$, $\log g$, and $\mathrm{[Fe/H]}$, respectively. Each regressor takes the original survey values of $T_{\rm eff}$, $\log g$, $\mathrm{[Fe/H]}$, and SNR as inputs, allowing the calibration to account for nonlinear dependencies among these quantities. Each MLP contains five hidden layers with 128, 64, 32, 16, and 8 neurons and rectified linear unit activations. The common sources are divided into training, validation, and test subsets in an approximate ratio of 70:15:15. Inputs and targets are standardized using the training subset. We use the Adam optimizer to minimize a squared-error loss weighted by the inverse variance of the APOGEE target parameter, with $L_2$ regularization and early stopping based on the validation loss.

Because Gaia-ESO and SEGUE have small samples in common with APOGEE, we use polynomial regressions to calibrate their giant and dwarf groups separately. For $T_{\rm eff}$ and $\log g$, we adopt five-term polynomials with cubic dependence on the parameter being calibrated and linear dependence on SNR:
\begin{equation}
T_{\rm eff, APOGEE} = a_0 + a_1 X_1 + a_2 X_1^2 + a_3 X_1^3 + a_4 Y,
\end{equation}

\begin{equation}
\log g_{\rm APOGEE} = b_0 + b_1 X_2 + b_2 X_2^2 + b_3 X_2^3 + b_4 Y,
\end{equation}
where $X_1$, $X_2$, and $Y$ represent the original $T_{\rm eff}$, $\log g$, and $\mathrm{SNR}$ values of the survey being calibrated. The coefficients $a_i$ and $b_i$ are fitted separately for each survey and stellar group. For metallicity in these two surveys, we fit an eight-term polynomial separately for giants and dwarfs:
\begin{equation}
[{\rm Fe/H}]_{\rm APOGEE} = c_0 + c_1 X_1 + c_2 X_1^2 + c_3 X_2 + c_4 X_2^2 + c_5 X_3 + c_6 X_3^2 + c_7 Y,
\end{equation}
where $X_1$, $X_2$, $X_3$, and $Y$ correspond to the original $T_{\rm eff}$, $\log g$, $\mathrm{[Fe/H]}$, and SNR, respectively. The coefficients are fitted separately for each survey and stellar group, using inverse-variance weights based on the APOGEE target uncertainties.

\begin{figure}[!htbp]
\centering
\includegraphics[width=0.98\textwidth,height=0.64\textheight,keepaspectratio]{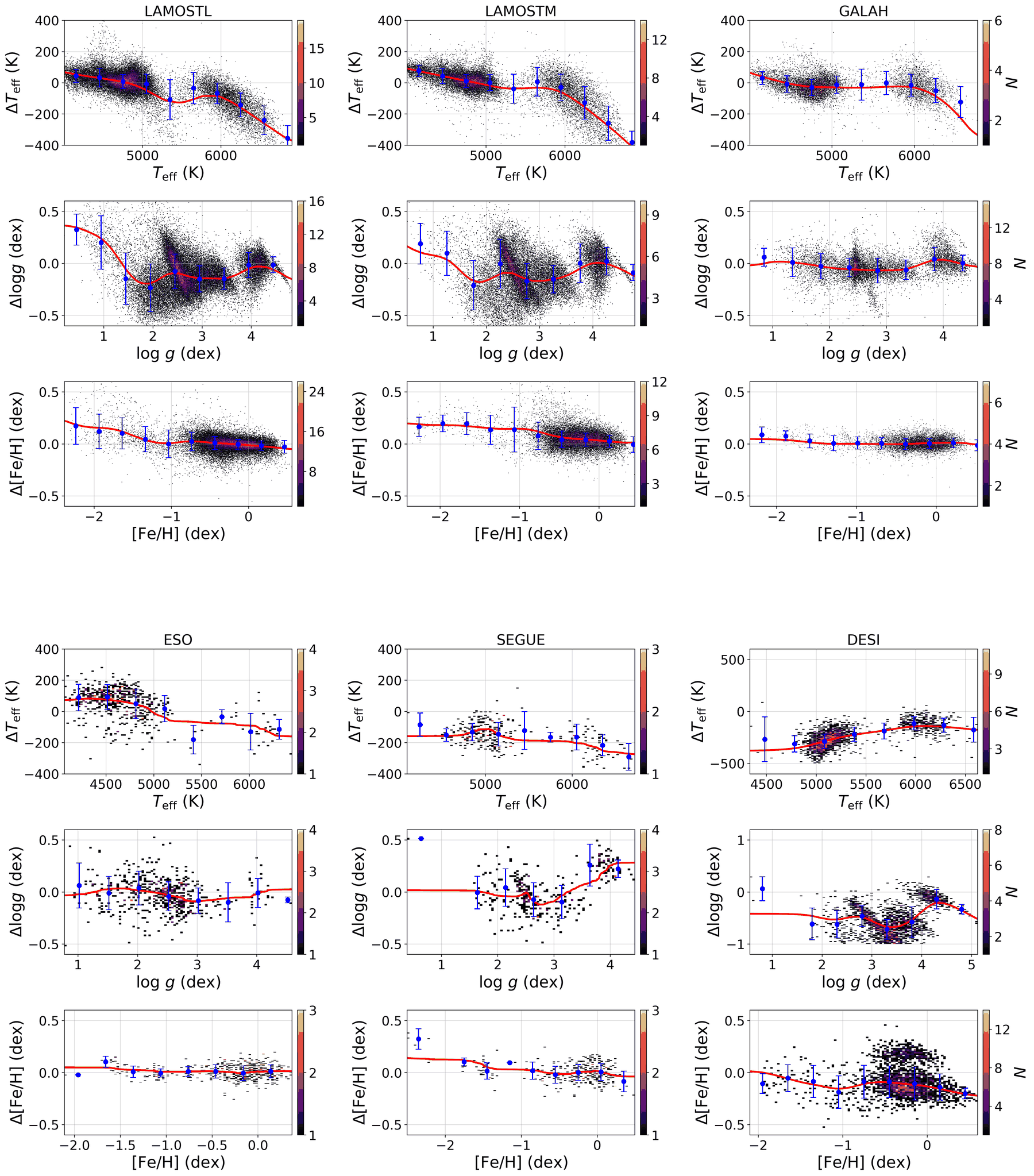}
\caption{Comparison of stellar atmospheric parameters between APOGEE and six complementary spectroscopic datasets (LAMOST LRS, LAMOST MRS, GALAH, Gaia-ESO, SEGUE, and DESI) for their common sources. For each survey panel, rows from top to bottom display results for $T_{\rm eff}$, $\log g$, and $\rm [Fe/H]$, respectively. Horizontal axes show the original parameters from each survey, and vertical axes show the differences in the sense of APOGEE minus the individual survey. Data points are color-coded by number density. Solid red lines show the predicted corrections from the separate giant and dwarf MLP or polynomial models. Blue symbols and error bars show the median observed differences and standard deviations within bins.}
\label{fig:multivariate polynomial regressions}
\end{figure}

Figure~\ref{fig:multivariate polynomial regressions} illustrates the cross-calibration results obtained from the common sources shared between APOGEE and the other individual surveys. Ordered from left to right and top to bottom, the panels display the original parameter differences and model-predicted correction trends for LAMOST LRS, LAMOST MRS, GALAH, Gaia-ESO, SEGUE, and DESI, respectively. The solid red lines summarize the model-predicted corrections, $p_{\rm corr}-p_{\rm original}$, as a function of the corresponding original parameter. The blue symbols and error bars show the median observed differences and their standard deviations within each bin. The LAMOST LRS and MRS temperature corrections follow the changes in the observed offset trends around $T_{\rm eff}\sim6000$\,K. The DESI comparisons also show nonlinear systematic differences, particularly in $\log g$, which are modeled using the separate giant and dwarf MLPs.

To construct our final science-ready spectroscopic sample, we relax the parameter uncertainty constraints to a broader yet acceptable threshold, requiring $\sigma_{T_{\rm eff}} < 200$\,K, $\sigma_{\log g} < 0.30$\,dex, and $\sigma_{\rm [Fe/H]} < 0.20$\,dex. We apply the corresponding MLP or polynomial calibrations to place the complementary catalogs on the APOGEE parameter scale. After systematically purging cross-survey duplicate entries, we obtain a combined sample consisting of 9,862,056 unique stars. This sample comprises 4,789,545 stars from LAMOST LRS, 1,526,750 from LAMOST MRS, 508,094 from APOGEE, 769,004 from GALAH DR4, 73,309 from Gaia-ESO DR5, 218,252 from SEGUE DR12, and 1,977,102 from DESI DR1. This value-added catalog provides the spectroscopic basis for our precise interstellar extinction estimates and can also serve as a training dataset for future data-driven models and automated stellar parameter estimation \citep{huang2022beyond, sun2023spar, sun2025photometric}.

\begin{figure}[!htbp]
\centering
\includegraphics[width=0.98\textwidth]{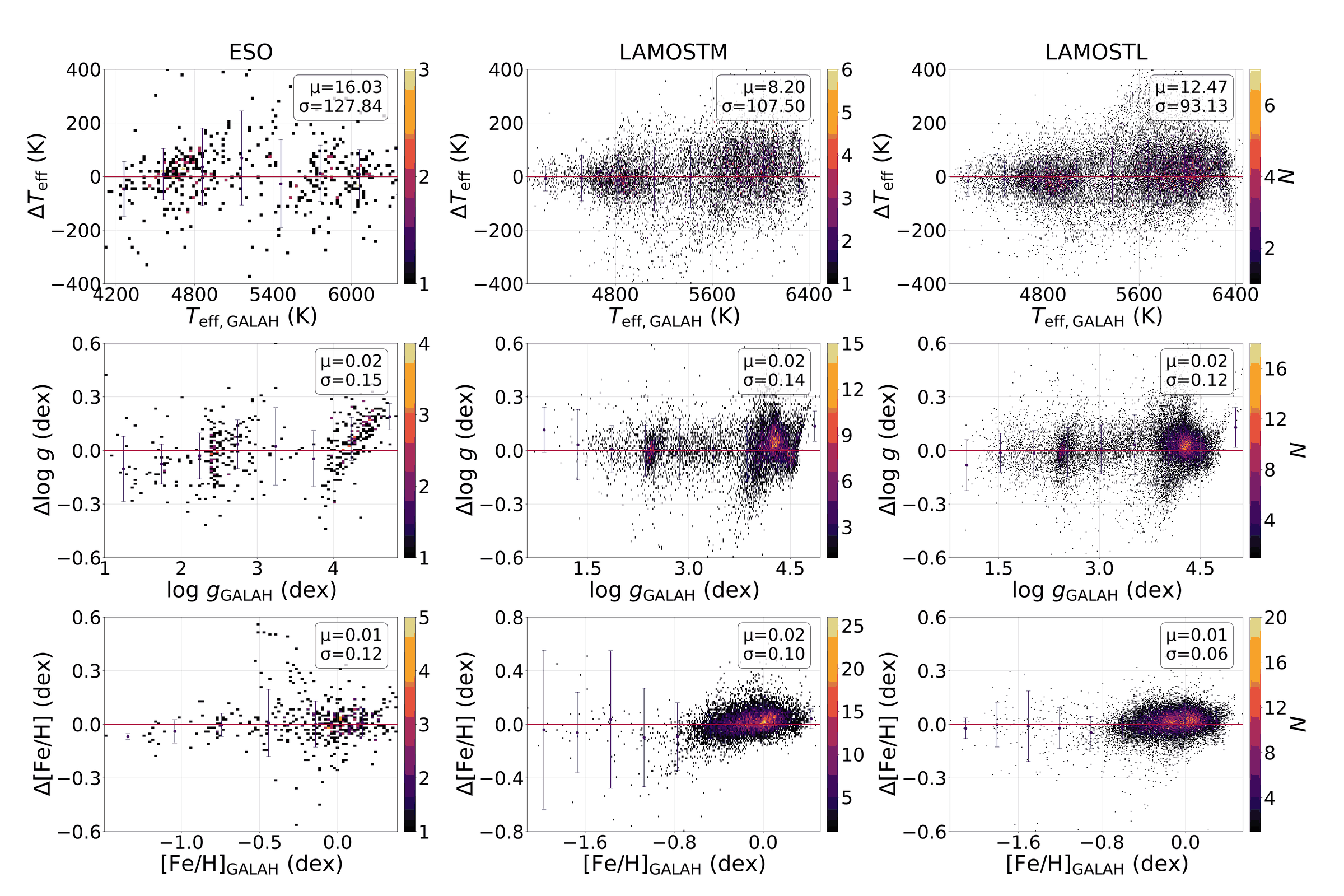}
\caption{Comparison of stellar atmospheric parameters between calibrated GALAH and three complementary spectroscopic datasets (Gaia-ESO, LAMOST MRS, and LAMOST LRS) for their common sources. Horizontal axes represent reference parameters from calibrated GALAH, while vertical axes show parameter residuals (individual survey minus GALAH). For each survey panel, rows from top to bottom display results for $T_{\rm eff}$, $\log g$, and $\rm [Fe/H]$, respectively. Data points are color-coded by number density. Blue symbols and error bars show the binned residuals and their dispersions. Red horizontal lines indicate zero difference. The annotated $\mu$ and $\sigma$ give the overall offset and dispersion for each comparison.}
\label{fig:galah_param_comparison}
\end{figure}

\begin{figure}[!htbp]
\centering
\includegraphics[width=0.98\textwidth]{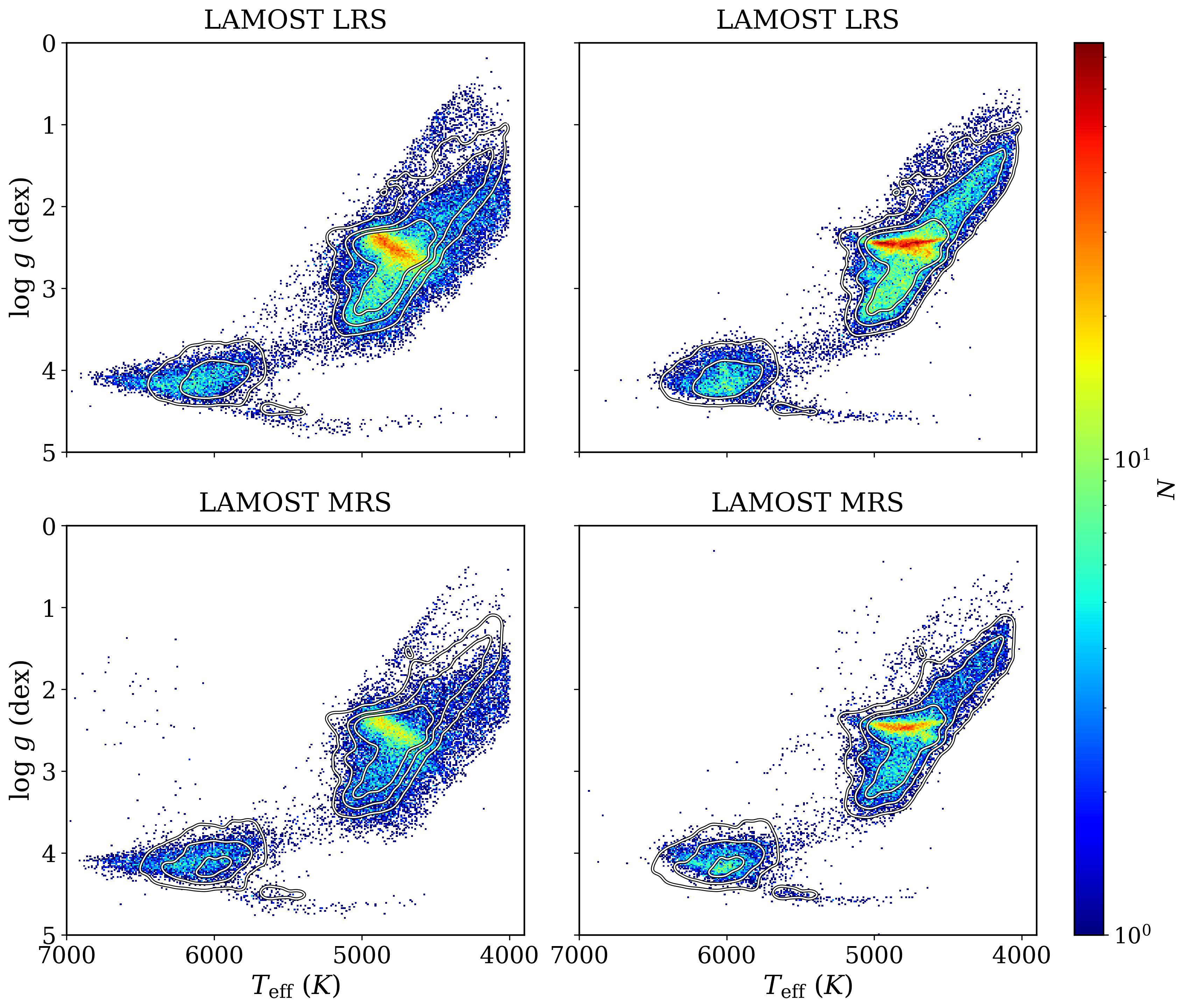}
\caption{The $T_{\rm eff}$--$\log g$ distributions of stars in common between LAMOST and APOGEE. The upper and lower rows show LAMOST LRS and LAMOST MRS, respectively. The left and right columns show the original and calibrated LAMOST parameters, respectively. The contours show the APOGEE distribution of the common sources.}
\label{fig:lamost_kiel_comparison}
\end{figure}

To assess consistency among the calibrated catalogs, we compare the corrected GALAH parameters with those of common sources in Gaia-ESO, LAMOST MRS, and LAMOST LRS, shown from left to right in Figure~\ref{fig:galah_param_comparison}. The residual distributions are generally concentrated near zero. The $\log g$ dispersions are 0.15, 0.14, and 0.12\,dex for Gaia-ESO, LAMOST MRS, and LAMOST LRS, respectively. A trend with $\log g$ remains in the Gaia-ESO comparison. We further examine the calibration results in the $T_{\rm eff}$--$\log g$ diagram. Figure~\ref{fig:lamost_kiel_comparison} compares the LAMOST LRS and MRS distributions before and after calibration, with APOGEE density contours overlaid for the common sources. After calibration, the giant branches follow the APOGEE contours more closely. The red-clump region becomes more concentrated around $\log g\simeq2.4$ and better aligned with the corresponding APOGEE density structure. The main dwarf locus also lies closer to the APOGEE contours. These changes are visible in both LAMOST LRS and MRS, indicating improved consistency with the APOGEE parameter scale.

We compare our spectroscopic parameters against a high-resolution benchmark compilation from the literature, the PASTEL database \citep{soubiran2016pastel}. The resulting median offsets and median absolute deviations (in the sense of our spectroscopic data minus PASTEL) are: $\Delta T_{\rm eff} = 9 \pm 141~{\rm K}$, $\Delta \log g = -0.01 \pm 0.21~{\rm dex}$, and $\Delta {\rm [Fe/H]} = 0.00 \pm 0.12~{\rm dex}$, indicating good overall agreement. To support potential future work on deriving stellar atmospheric parameters from photometric data, we also evaluate the metallicity distributions of member stars in globular clusters, as summarized in Table~\ref{tab:globular clusters}. The second column lists reference values provided by \citet{helmi2018gaia}, while the third and fourth columns display mean values and dispersions derived from member stars in our spectroscopic sample. The cluster mean metallicities differ from the reference values by no more than 0.11\,dex. However, the dispersion exhibits a noticeable increase toward lower metallicities.

\begin{table}
\bc
\begin{minipage}[]{100mm}
\caption[]{Properties and metallicity measurements of globular clusters compiled in our spectroscopic sample. The columns list the cluster name, the reference $\text{[Fe/H]}_{\text{H18}}$ from \citet{helmi2018gaia}, and mean metallicity ($\langle\text{[Fe/H]}\rangle$) along with its standard deviation ($\sigma_{\text{[Fe/H]}}$) derived from member stars in our spectroscopic sample.
\label{tab:globular clusters}}
\end{minipage} \\
\setlength{\tabcolsep}{2.5pt}
\small
\begin{tabular}{cccc}
\hline\noalign{\smallskip}
Cluster & $\rm [Fe/H]_{\rm H18}$ (dex) & $\langle\rm [Fe/H]\rangle$ (dex) & $\sigma_{\rm [Fe/H]}$ (dex) \\
\hline\noalign{\smallskip}
NGC104  & -0.72 & -0.76 & 0.05 \\
NGC6656 & -1.70 & -1.69 & 0.13 \\
NGC6121 & -1.16 & -1.06 & 0.07 \\
NGC5139 & -1.53 & -1.61 & 0.23 \\
NGC6752 & -1.54 & -1.50 & 0.06 \\
NGC5904 & -1.29 & -1.22 & 0.08 \\
NGC5272 & -1.50 & -1.43 & 0.15 \\
NGC6218 & -1.37 & -1.27 & 0.07 \\
\noalign{\smallskip}\hline
\end{tabular}
\ec
\end{table}

\subsection{The Star-Pair Algorithm} \label{subsec:the star-Pair algorithm}
The underlying premise of the Star-Pair algorithm is that stars sharing identical atmospheric parameters---$T_{\rm eff}$, $\log g$, and $\rm [Fe/H]$---possess the identical intrinsic color.
In this work, we use a combined control sample to increase the availability of suitable comparison stars across the target sample. Applying the method separately within each survey is also feasible when its control sample adequately covers the stellar-parameter distribution of its target stars. However, restricting the controls to individual surveys reduces both their number and their coverage of stellar-parameter space. A test of this approach yielded an overall matching success rate of 96.11\%. GALAH and SEGUE, with only 5,281 and 2,450 control stars, achieved matching success rates of 85.24\% and 61.08\%, respectively. Compared with using the combined control sample, the other surveys also showed decreases in matching success rate, although to a lesser extent. We therefore calibrate the atmospheric parameters to the APOGEE scale before combining the control samples, enabling targets from different surveys to share a larger pool of comparison stars. After parameter calibration, the $E(B-V)$ estimates from different surveys show slightly improved overall consistency.

To eliminate the artificial clustering effects caused by the disparate dimensional scales of these physical quantities, the three-dimensional parameter space is initially rescaled using highly customized anisotropic weights (\texttt{[100.0, 0.2, 0.1]}). A high-dimensional spatial index is subsequently established via a \texttt{cKDTree} framework. Utilizing this index, the algorithm executes a deeply nested, four-tier adaptive fitting decision flow for each target star as detailed below:
\begin{enumerate}
    \item \textbf{Dense Enclosure Zone Identification}: When the number of neighboring control stars $N$ within a localized radius $r \le 1.0$ satisfies $N \ge 10$, the program verifies whether these neighbors form a valid geometric enclosure (\textit{Surround}) enclosing the target source across the parameter axes. If this spatial criterion is met, the domain is deemed a secure interpolation regime. The algorithm then extracts the 10 closest stars and applies an Inverse Distance Weighting (IDW) mean to determine the predicted intrinsic color.
    \item \textbf{Dense Boundary Zone Degradation}: If $N \geq 10$ but the neighbors fail to achieve the required spatial enclosure, the target source is identified as residing on the peripheral boundary of the control sample, thereby posing an extrapolation risk. In this scenario, the algorithm adaptively degrades its constraints, expands the search radius to $r = 1.5$ to retrieve a larger sample of neighbors for a broader IDW calculation.
    \item \textbf{Sparse Transition Zone Processing}: When the local neighbor count within $r \leq 1.0$ falls into a sparse regime ($3 \leq N < 10$), the spatial enclosure test is bypassed. The algorithm directly utilizes all available valid neighbors within this local spherical domain to perform a standard localized IDW estimation.
    \item \textbf{Extreme Sparsity and Edge Extrapolation}: In cases of extreme localized starvation where $1 \leq N < 3$, the algorithm forcibly expands its spatial boundaries. If at least 3 control stars can be secured within an extended radius of $r = 1.5$, the IDW method is retained. Otherwise, the ultimate extrapolation mechanism is triggered: the algorithm selects the 20 nearest control stars unconditionally from the entire library and introduces a high-order, non-linear Radial Basis Function (\texttt{Rbf}) with a \texttt{multiquadric} kernel to perform a multi-dimensional surface interpolation.
\end{enumerate}

Following the successful prediction of the target source's expected intrinsic color $(G_{\mathrm{BP}} - G_{\mathrm{RP}})_0$ via the adaptive decision flow, the color excess $E(BP - RP)$ is directly determined by subtracting the predicted intrinsic color from the observed apparent color $(G_{\mathrm{BP}} - G_{\mathrm{RP}})$. To ensure the statistical validity and reliability of the derived extinction measurements, the total uncertainty associated with $E(BP - RP)$ is calculated by accounting for three independent physical error contributors, modeled via a geometric root-sum-square aggregation:

\begin{equation}
    \sigma_{E(BP-RP)} = \sqrt{\sigma_{\mathrm{phot}}^2 + \sigma_{\mathrm{param}}^2 + \sigma_{\mathrm{fit}}^2}
\end{equation}

where $\sigma_{E(BP-RP)}$ represents the total uncertainty of the color excess $E(BP - RP)$, $\sigma_{\mathrm{phot}}$ denotes the combined photometric uncertainty from $G_{\mathrm{BP}}$ and $G_{\mathrm{RP}}$ bands, $\sigma_{\mathrm{param}}$ represents the error propagated from the stellar atmospheric parameters, and $\sigma_{\mathrm{fit}}$ is the intrinsic fitting error derived from the star-pair algorithm.

\begin{figure}
\centering
\includegraphics[width=0.45\textwidth]{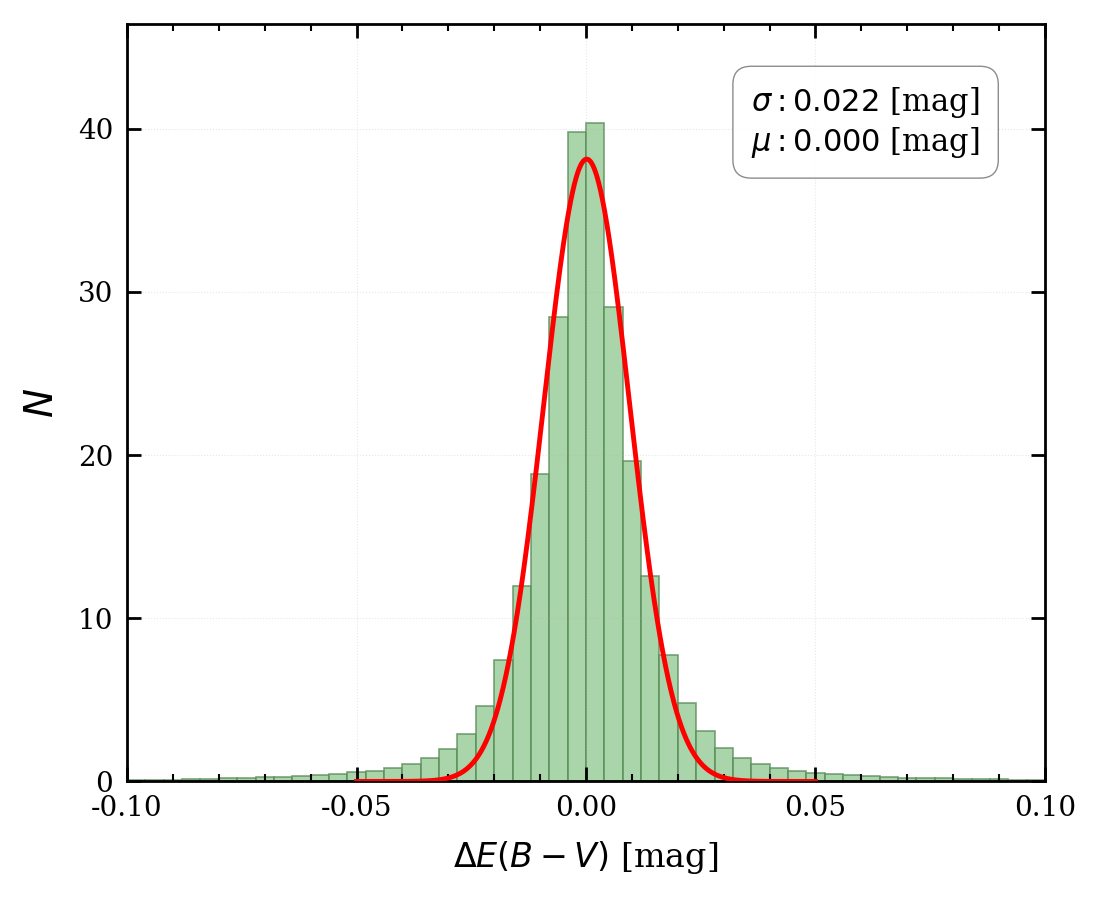}
\caption{Distribution of $E(B-V)$ residual differences derived from independent multi-epoch measurements of LAMOST duplicate sources using the star-pair method. The red solid line represents the best-fit Gaussian distribution.}
\label{fig:lamost_duplicate_ebv}
\end{figure}

To quantitatively evaluate the precision and internal consistency of the star-pair algorithm, we cross-matched our sample with the LAMOST Low-Resolution Spectroscopic (LRS) Multiple Epoch Catalogue against the control sample. Figure~\ref{fig:lamost_duplicate_ebv} illustrates the resulting distributions of the mean ($\mu$) and standard deviation ($\sigma$) of color excesses derived for individual multi-epoch stars. Fitting this residual distribution with a Gaussian profile yields a characteristic dispersion of $\sigma \approx 0.022~{\rm mag}$, which directly quantifies the intrinsic extinction measurement precision achieved by our algorithm.

\section{Precise Interstellar Extinction Estimates for Nearly Ten Million Stars} \label{sec:precise extinction estimates}

\subsection{The catalog} \label{subsec:the catalog}

Based on the star-pair method, we first cross-matched our spectroscopic sample with the Gaia EDR3 catalog \citep{brown2021gaia}, applying a stringent quality criterion that retains only sources with photometric uncertainties in both $G_{\rm BP}$ and $G_{\rm RP}$ bands of less than $0.03~{\rm mag}$. Following previous methodologies \citep[e.g.,][]{yuan2013empirical, chen2024photometric, wang2025all}, a control sample of $286,447$ stars was constructed from regions with low foreground dust where the \citet[SFD;][]{schlegel1998maps} reddening values satisfy $E(B-V) < 0.01$\,mag, leaving the remaining $9,575,609$ sources as the target sample. For stars in the control sample, we applied a scale factor of $0.86$ to the SFD reddening values to correct for the systematic difference relative to stellar locus measurements \citep{schlafly2010blue, schlafly2011measuring, yuan2013empirical}. By applying the star-pair matching algorithm, a total of $9,329,941$ stars within the target sample were successfully matched, achieving a high success rate of $97.43\%$. Combining the control sample with these successfully matched target stars yielded a final dataset of $9,616,388$ stars, which has been publicly released at \url{https://nadc.china-vo.org/res/r101883/}. A comprehensive description of this final catalog is provided in Table~\ref{tab:extinction_catalog_description}; it delivers essential stellar parameters—including equatorial coordinates (RA, DEC), Galactic coordinates ($l$, $b$), and distances derived from either \citet{zhang2025three} or \citet{bailer2021estimating}—along with the $E(BP-RP)$ color excesses and uncertainties calculated via the star-pair technique, and the $A_v$ absolute extinction and $E(B-V)$ reddening values corrected based on the extinction law of \citet{zhang2023empirical}.

To facilitate flexible access and practical applications of this massive dataset, a standalone Python pipeline is integrated into the online repository alongside the catalog. This standalone pipeline allows users to dynamically query and extract the interstellar extinction $E(B-V)$ for any custom target coordinates $(l, b)$ and specified distance $d$, utilizing either a single HEALPix pixel \citep{gorski2005healpix} or a user-defined cone search radius. Upon selecting the local stellar sample around the specified target direction, the pipeline executes a robust dual-stage optimization scheme to derive the line-of-sight extinction profile, as detailed in Section~\ref{subsec:calculation of Line-of-Sight extinction profiles}.

\begin{table}
\bc
\begin{minipage}[]{100mm}
\caption[]{Description of the interstellar extinction catalog.\label{tab:extinction_catalog_description}}\end{minipage} \\
\setlength{\tabcolsep}{2.5pt}
\small
 \begin{tabular}{cclc}
  \hline\noalign{\smallskip}
{Column} & {Label} & {Description} & {Unit} \\
  \hline\noalign{\smallskip}
1 & source\_id & Unique source identifier for Gaia DR3 \\ 
2 & RA & Right Ascension from Gaia DR3 & degrees \\
3 & DEC & Declination from Gaia DR3 & degrees \\
4 & $l$ & Galactic longitude & degrees \\
5 & $b$ & Galactic latitude & degrees \\
6 & d & Distance from \citet{zhang2025three} or \citet{bailer2021estimating} & kpc \\
7 & d\_err & Error of distance from \citet{zhang2025three} or \citet{bailer2021estimating} & kpc \\
8 & EBPRP & $E(BP-RP)$ from this work & mag \\
9 & EBPRP\_err & Error of $E(BP-RP)$ from this work & mag \\
10 & EBV & E(B-V) from this work & mag \\
11 & EBV\_err & Error of E(B-V) from this work & mag \\
12 & Av & $A_{\rm v}$ from this work & mag \\
13 & Av\_err & Error of $A_{\rm v}$ from this work & mag \\
  \noalign{\smallskip}\hline
\end{tabular}
\ec
\end{table}

\subsection{Calculation of Line-of-Sight Extinction Profiles}
\label{subsec:calculation of Line-of-Sight extinction profiles}

To derive line-of-sight reddening profiles, we model the cumulative colour excess $E(B-V)$ along each line of sight as a monotonically increasing step function \citep[e.g.,][]{zhang2026gsed}.
For a distance $d$ within the $k$th distance bin, the cumulative reddening $E(d)$ is defined as:
\begin{equation}
    E(d) = \sum_{i=1}^{k} e^{h_i},
    \quad \text{where } d \in \text{bin}_k
\end{equation}

Here, the exponential form $e^{h_i}$ ensures that each incremental extinction $\Delta E_i$ is strictly positive, physically guaranteeing that $E(d)$ grows monotonically.
We employ a Markov Chain Monte Carlo (MCMC) approach to explore the posterior distribution $P(\theta|{\rm Data}) \propto \mathcal{L}({\rm Data}|\theta) \cdot P(\theta)$.
To enhance robustness against outliers, we adopt a log-likelihood function inspired by the Cauchy distribution:
\begin{equation}
    \ln\mathcal{L}
    = -\sum_{i=1}^{N}
    \ln \left(
        1 +
        \left[
            \frac{E_i-f(d_i,\theta)}{\sigma_{E_i}}
        \right]^2
    \right)
\end{equation}
Here, $E_i$ and $d_i$ are the observed reddening and adopted distance of star $i$, and $f(d_i,\theta)$ is the model prediction. Following \citet{chen2019three}, we account approximately for distance uncertainties through the effective reddening uncertainty:
\begin{equation}
    \mbox{$\displaystyle
    \sigma_{E_i}
    =
    \sqrt{
        \sigma_{{\rm obs},i}^{2}
        +
        \left(
            E_i\frac{\sigma_{d_i}}{d_i}
        \right)^{2}
    }$}
    \label{eq:los_combined_uncertainty}
\end{equation}
Here, $\sigma_{{\rm obs},i}$ and $\sigma_{d_i}$ are the reddening measurement uncertainty and distance uncertainty, respectively. This prescription assumes approximately uniform dust opacity along the line of sight. The combined uncertainties are used in both the preliminary least-squares fit and the MCMC likelihood, while stars remain assigned to bins according to their adopted distances.

This logarithmic penalty effectively reduces the weight of large residuals compared to standard $\mathcal{L}_2$ norms.
A preliminary nonlinear least-squares fit provides starting values, which are perturbed to initialize the MCMC chains. Each distance bin is represented by a single reddening level, whereas the stellar reddening can vary within that bin. Where reddening increases rapidly, a fitted step can therefore lie above the nearer stars and below the more distant stars within the same bin. Together with the monotonicity constraint and the uncertainty-weighted Cauchy-inspired likelihood, this explains why the fitted profile need not pass through every local concentration of points in the scatter plot.

To facilitate community-driven investigations and reproducibility, the complete Python code implementing this fitting pipeline has been made publicly available.
This open-source tool enables users to extract the interstellar reddening $E(B-V)$ for a requested distance within the fitted range along any custom-defined line of sight.
Within this framework, users are granted the flexibility to either specify a customized cone search radius around the target coordinates to dynamically adjust the sample density, or simply proceed with the default HEALPix pixel.
Upon identifying the stellar sample within the designated spatial volume, the algorithm executes the dual-stage optimization detailed above, outputting $E(B-V)$ estimates dynamically tailored to the user's input distance parameter.
Specifically, the pipeline automatically filters unphysical data, applies a quantile-based robust mask to prevent leverage from extreme outliers, and dynamically determines the distance binning according to the local stellar density.
The algorithm then executes the MCMC optimization to evaluate $E(B-V)$, generating and saving high-resolution extinction profile plots with labeled target parameters automatically.

\begin{figure}
\centering
\includegraphics[width=0.98\textwidth]{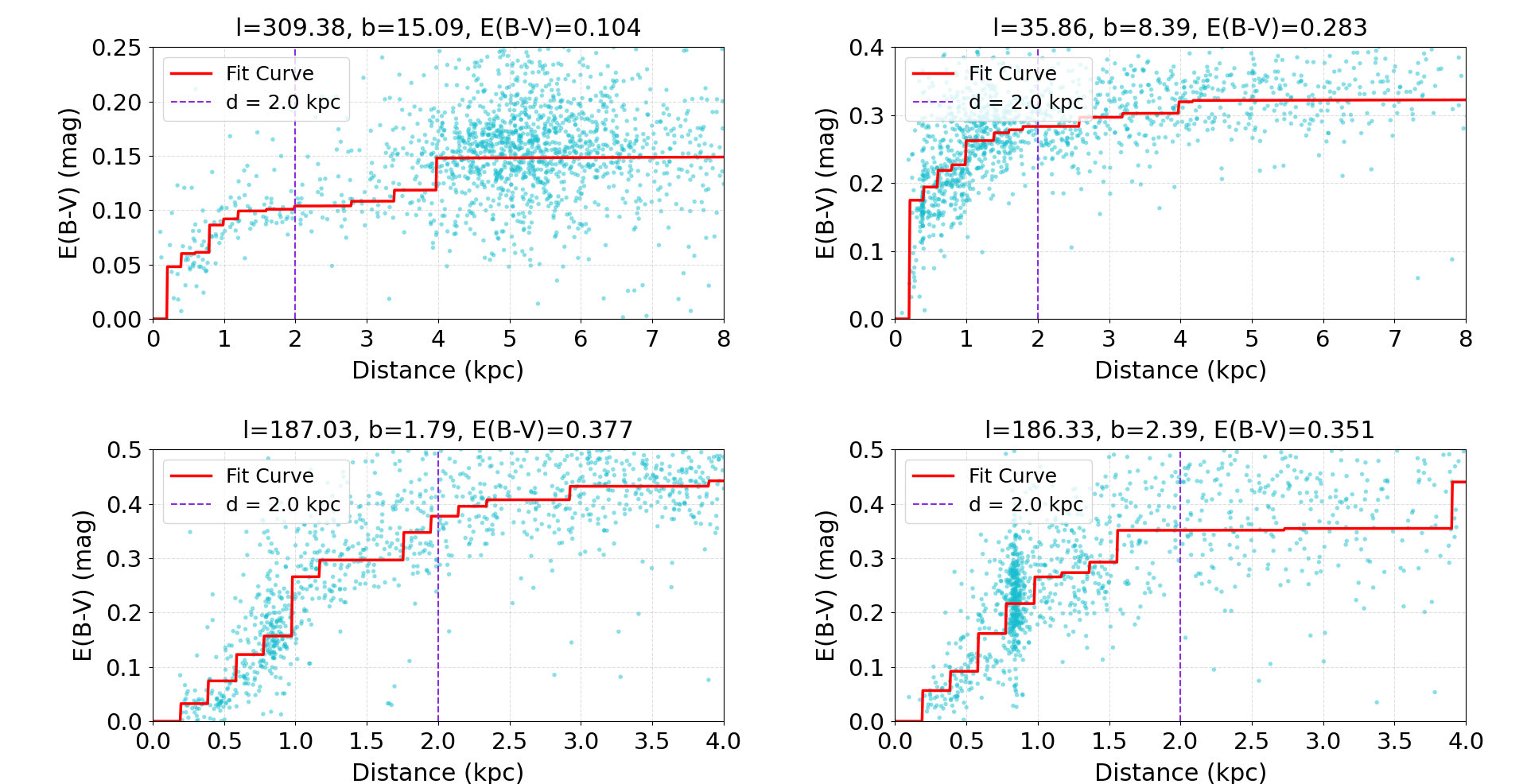}
\caption{$E(B-V)$ as a function of distance for the four HEALPix pixels containing the largest numbers of stars. For each panel, the Galactic coordinates $(l, b)$ of the pixel center and the resulting modeled $E(B-V)$ are indicated at the top. Cyan dots represent individual stars with their respective distances and star-pair derived color excesses. The solid red curves illustrate our fitted cumulative reddening profiles, while the vertical dashed purple lines mark the requested distance of $2.0~{\rm kpc}$.}
\label{fig:ebv_distance_profiles}
\end{figure}

Figure~\ref{fig:ebv_distance_profiles} illustrates representative examples of this fitting pipeline applied to the four HEALPix pixels with the highest stellar densities.
The fitted cumulative extinction profiles (solid red curves) show the overall increase in reddening with distance, while the individual stars (cyan dots) show the scatter in the stellar reddening measurements.
For a given input distance of $d = 2.0~{\rm kpc}$ (indicated by the vertical dashed purple lines), the derived $E(B-V)$ reddening values for these four specific regions are $0.104~{\rm mag}$, $0.283~{\rm mag}$, $0.377~{\rm mag}$, and $0.351~{\rm mag}$, respectively, in order from the upper-left to the lower-right panel.
This demonstrates the application of the fitting pipeline across diverse lines of sight with differing dust columns.

\subsection{Three-dimensional Extinction Maps} \label{subsec:3D Extinction Maps}

To characterize the spatial tomography and large-scale distribution of interstellar dust, we project our final catalog into a series of distance intervals. For each individual distance bin, stars falling within the designated distance window are grouped into spatial pixels using a HEALPix resolution of $N_{\rm side} = 64$ ($\sim0.839\text{~deg}^2$). This specific resolution is adopted to balance spatial detail with statistical robustness; across the entire grid, pixels containing fewer than 30 stars account for only $11.96\%$ of the total sky, ensuring a highly complete sampling framework. Within each active pixel, the characteristic $E(B-V)$ interstellar reddening value is explicitly determined by calculating the 95th percentile of the star-pair derived color excesses of the stars enclosed in that specific spatial-distance volume. Consequently, the final high-quality 3D reddening maps successfully cover a total comprehensive sky area of $33,894.87~{\rm deg}^2$.

\begin{figure}
\centering
\includegraphics[width=0.49\textwidth]{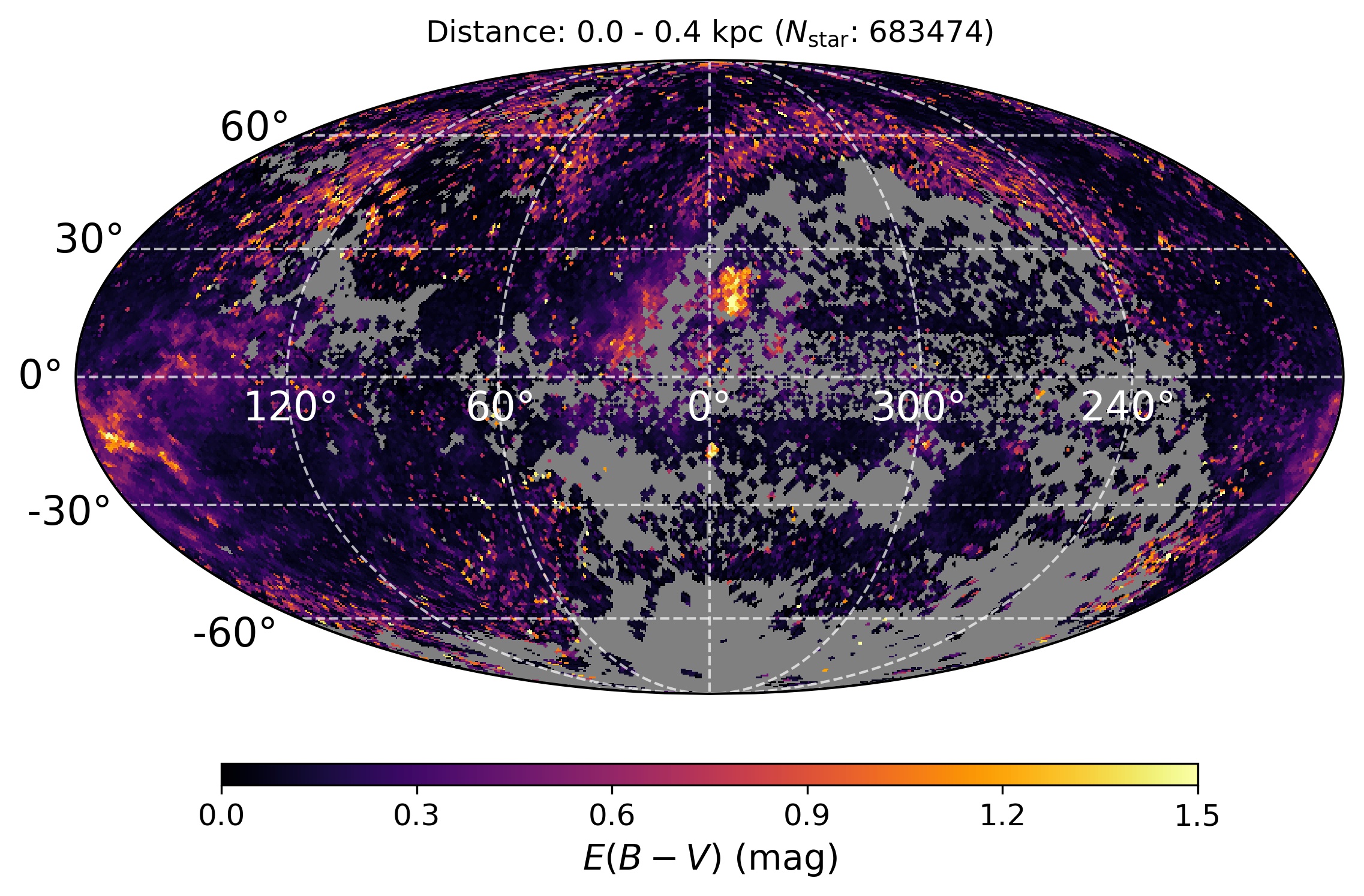}
\includegraphics[width=0.49\textwidth]{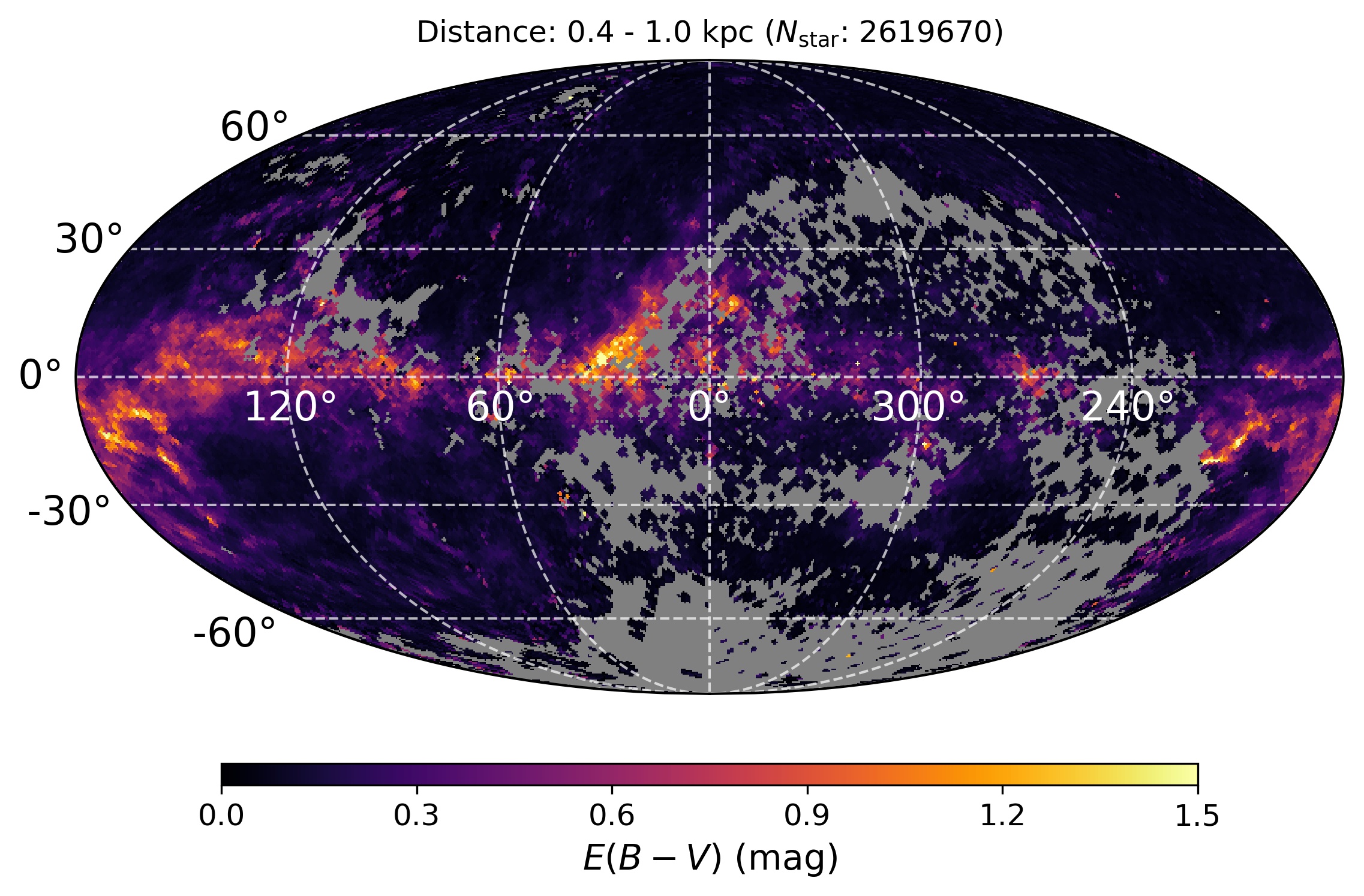}
\includegraphics[width=0.49\textwidth]{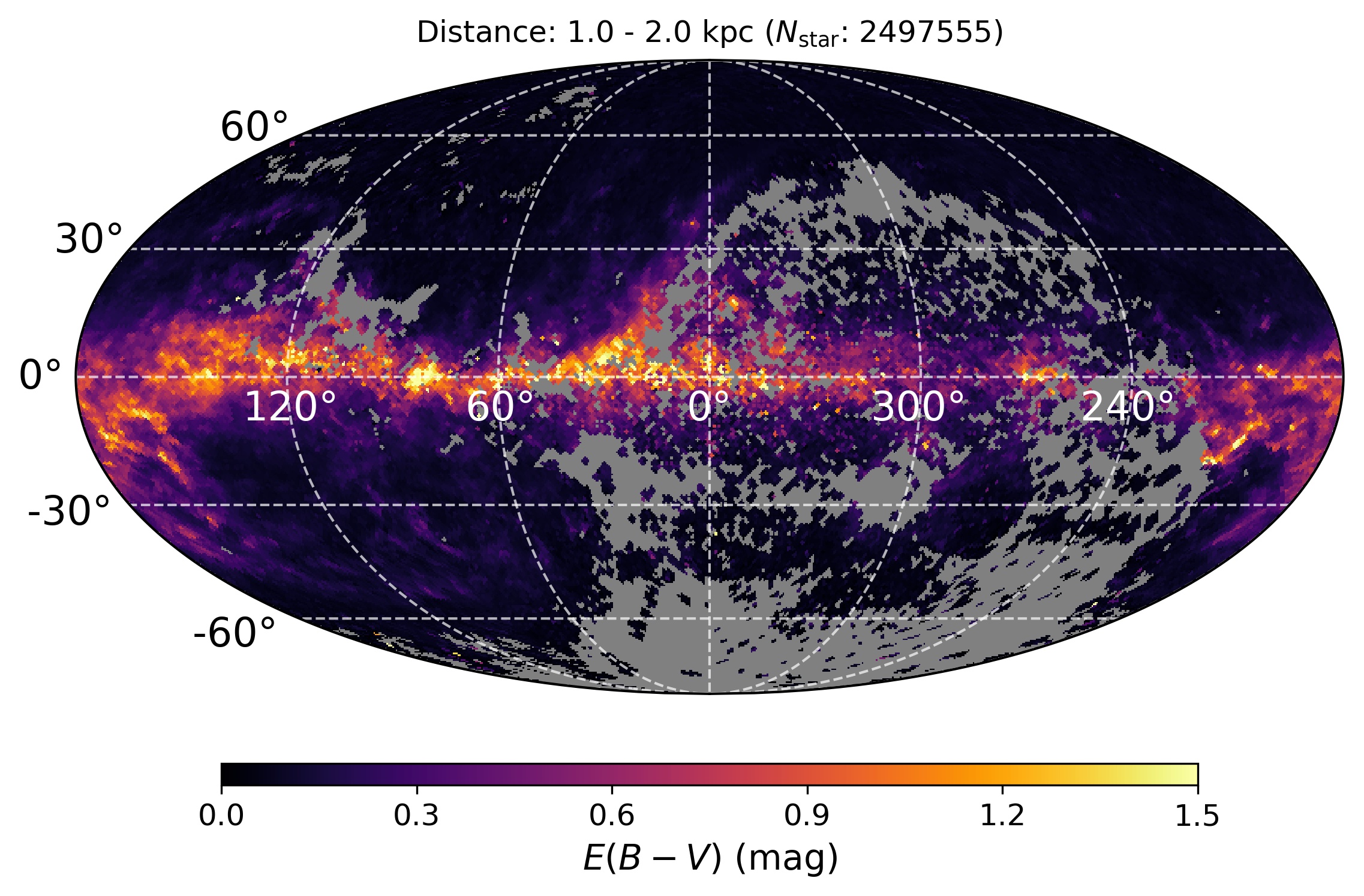}
\includegraphics[width=0.49\textwidth]{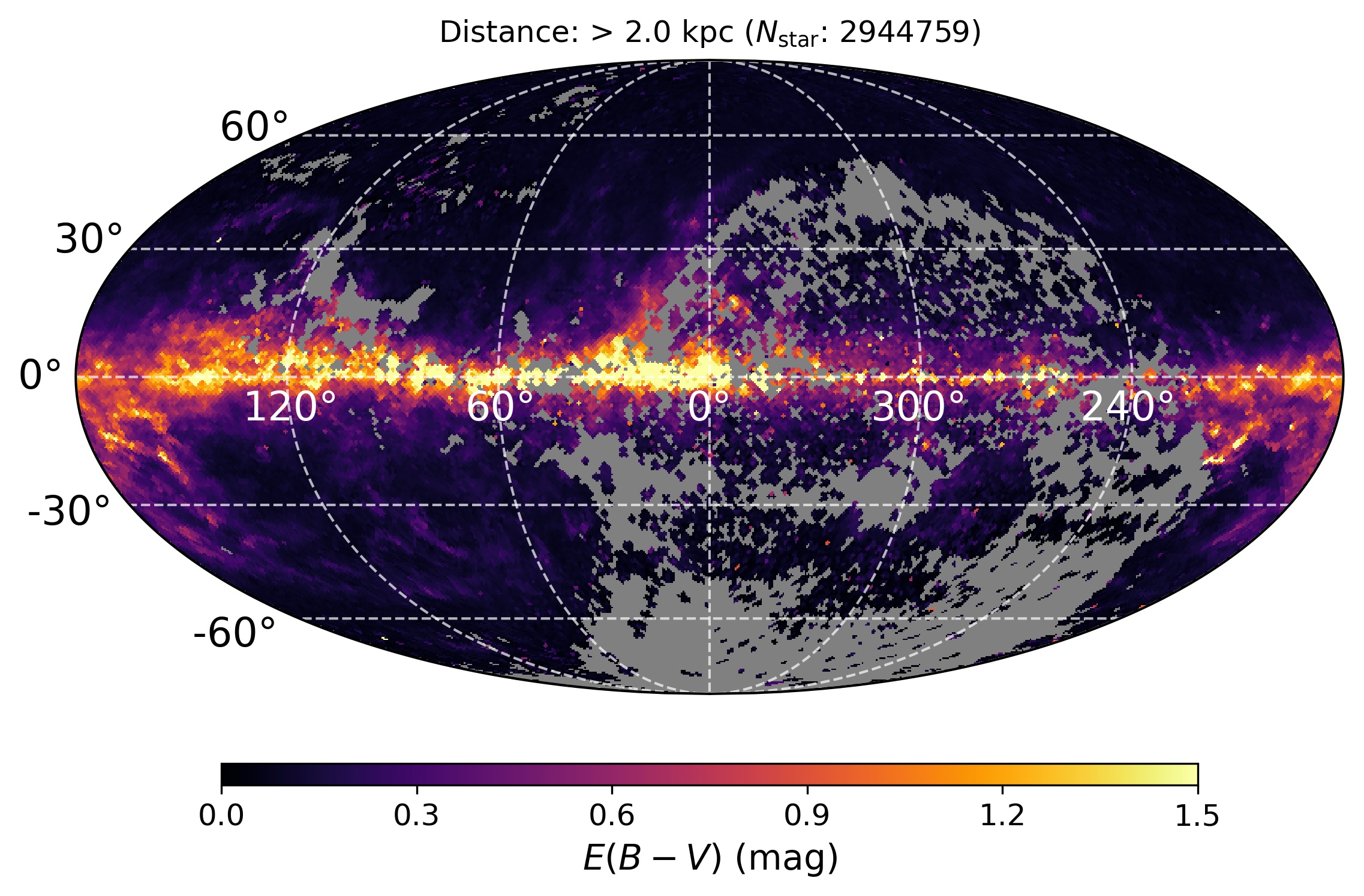}
\caption{Cumulative and bin-sliced $E(B-V)$ reddening sky maps shown in Galactic coordinates with a Mollweide projection ($l = 0^\circ$ centered). The panels display distributions for four discrete distance slices ($d < 0.4~{\rm kpc}$, $0.4 \le d < 1.0~{\rm kpc}$, $1.0 \le d < 2.0~{\rm kpc}$, and $d \ge 2.0~{\rm kpc}$). The distance range and star count ($N_{\mathrm{star}}$) are marked atop each panel. The color bar denotes $E(B-V)$ in mag, and grey shaded areas represent regions without sample coverage.}
\label{fig:ebv_3d_maps}
\end{figure}

Figure~\ref{fig:ebv_3d_maps} presents the cumulative and bin-sliced $E(B-V)$ extinction sky maps across different distance ranges derived from our catalog, mapped on a HEALPix grid in Galactic coordinates with a Mollweide projection, where a quality cut of $\sigma_d / d < 0.2$ was applied. At small distances ($d < 0.4~{\rm kpc}$), several well-known local dust cloud structures are clearly identifiable, particularly near the Galactic anti-center region, such as the Taurus and Perseus molecular clouds. As the distance increases to $0.4\text{--}1.0~{\rm kpc}$, prominent structures in the anti-center direction—including the Orion and Monoceros OB1 molecular cloud complexes—become increasingly conspicuous. At intermediate distances of $1\text{--}2~{\rm kpc}$, a series of major, highly opaque dust complexes situated along the Galactic plane dominate the sky maps, including W3, Cygnus X, Monoceros R2, and the extended Orion complex. At larger distances ($d \ge 2.0~{\rm kpc}$), the high-extinction dust distributions become strongly confined to a narrow strip along the Galactic plane. Overall, while localized dust structures are prevalent at nearby distances, the high-extinction regions at larger distances systematically concentrate tightly toward the Galactic plane, demonstrating excellent agreement with the known large-scale features of the interstellar dust distribution.

\section{Tests and Applications of Extinction Estimates} \label{Tests and Applications of Extinction Estimates}
To rigorously validate the precision and demonstrate the practical utility of our derived $E(B-V)$ reddening estimates, we perform comparative evaluations and illustrative applications. First, we cross-compare our catalog with four representative extinction studies in the literature \citep{wang2025all, zhang2025three, guo2021three, chen2019three}. Next, we demonstrate the effectiveness of our extinction corrections by examining the color magnitude diagram (CMD) of stars in two representative sky fields before and after dereddening.

\subsection{Comparison with Previous Extinction Studies} \label{subsec:comparison with previous extinction studies}

\begin{figure}
\centering
\includegraphics[width=0.98\textwidth]{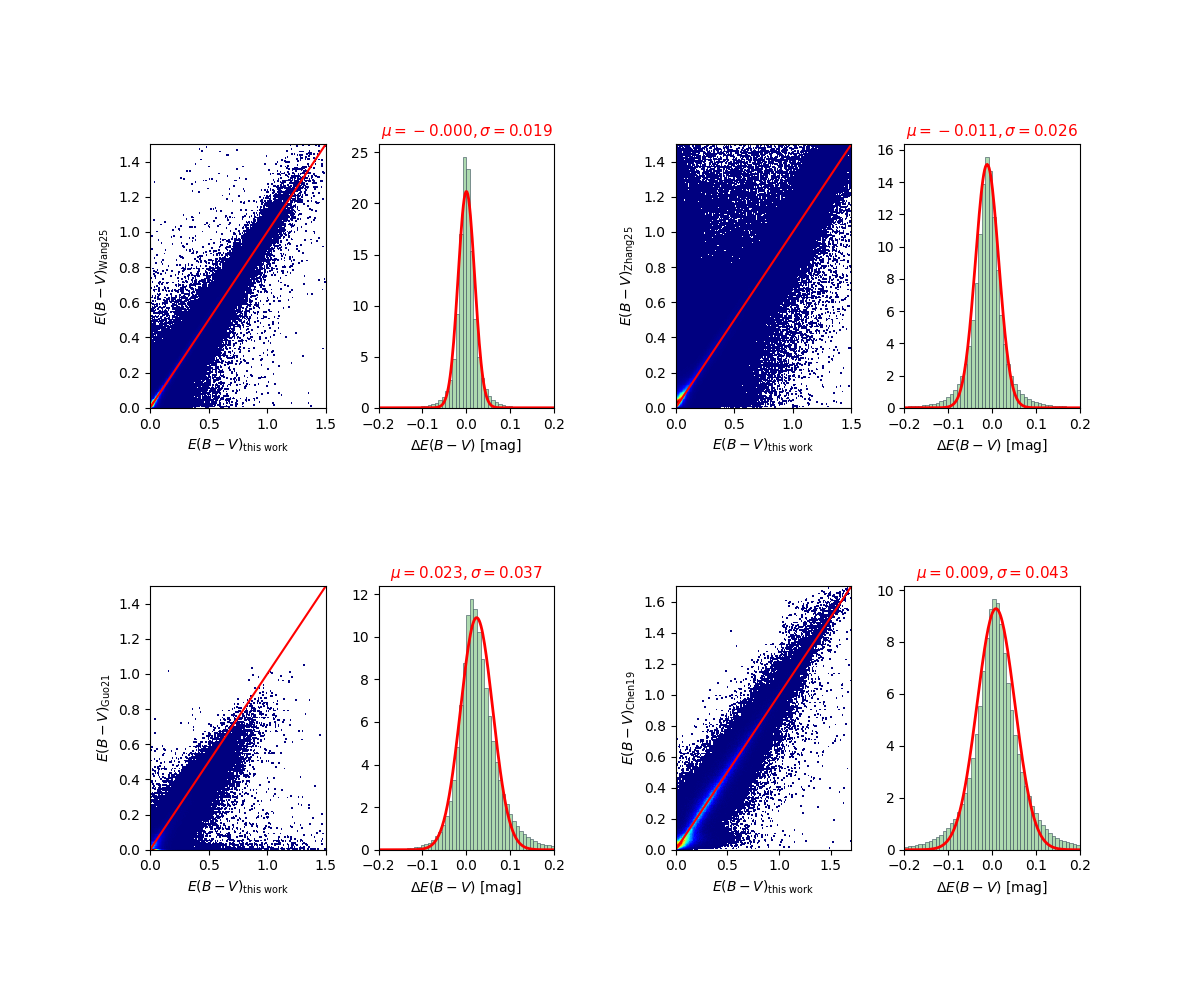}
\caption{Comparison of $E(B-V)$ values derived in the current work and those from \citet{wang2025all} (upper left), \citet{zhang2025three} (bottom left), \citet{guo2021three} (upper right) and \citet{chen2019three} (bottom right). Red straight lines denoting complete equality have been plotted as reference, while red curves are Gaussian fits of the distribution of differences between the maps. The corresponding means and standard deviations are on top of each histogram.}
\label{fig:compare_ebv_literature}
\end{figure}

The extinction estimates of individual stars derived in this work are compared with those from four representative studies \citep{wang2025all, zhang2025three, guo2021three, chen2019three} in Figure~\ref{fig:compare_ebv_literature}. Overall, our results exhibit excellent consistency with previous maps across diverse survey footprints.

For the star-pair benchmark map by \citet{wang2025all}, restricting the analysis to reliable estimates with $E(B-V)_{\rm err} < 0.10~{\rm mag}$ yields 2,874,197 common stars. The residual distribution ($E(B-V)_{\rm ours} - E(B-V)_{\rm Wang25}$) shows a negligible offset of $\mu = -0.000~{\rm mag}$ with a tight dispersion of $\sigma = 0.019~{\rm mag}$, demonstrating outstanding internal consistency.

Comparing with the Gaia XP forward-modeling map of \citet{zhang2025three} ($E(B-V)_{\rm err} < 0.15~{\rm mag}$) produces the largest overlap in our comparison, with 6,697,947 sources in common. The extinction differences exhibit a minor mean offset of $\mu = -0.011~{\rm mag}$ and a scatter of $\sigma = 0.026~{\rm mag}$. A small fraction of outliers appears in the upper-left scatter region where our $E(B-V)$ is small while theirs is overestimated ($E(B-V) > 0.8~{\rm mag}$). This discrepancy stems primarily from their inferred stellar parameters: for stars that are spectroscopically metal-poor (${\rm [Fe/H]} < 0$), \citet{zhang2025three} systematically overestimate metallicities (${\rm [Fe/H]} > 0$), propagating into artificially elevated reddening values.

For the SED-based southern map of \citet{guo2021three}, converting their $A_r$ values to $E(B-V)$ yields 656,461 stars in common. The comparison shows an average offset of $\mu = 0.015~{\rm mag}$ and an rms dispersion of $\sigma = 0.038~{\rm mag}$. The noticeable systematic offset in high-extinction regions ($E(B-V) > 0.3~{\rm mag}$) is likely caused by their stellar loci implementation, which accounts only for $T_{\rm eff}$ while $\log g$ and [Fe/H] dependencies on intrinsic colors.

Finally, for the machine-learning map of \citet{chen2019three}, enforcing ${\rm UEBP} - {\rm DEBR} < 0.15~{\rm mag}$ and adopting $E(B-V) = 0.75 \times E(G_{\rm BP} - G_{\rm RP})$ yields 850,030 common sources. The differences exhibit a mean offset of $\mu = -0.009~{\rm mag}$ and a dispersion of $\sigma = 0.043~{\rm mag}$. Direct comparisons using raw $E(G_{\rm BP} - G_{\rm RP})$ values yield virtually identical statistical performance, confirming the robustness of our intrinsic color determinations.

\subsection{Application to Extinction-Corrected Color--Magnitude Diagrams}
\label{subsec:cmd_application}

\begin{figure}
\centering
\includegraphics[width=0.98\textwidth]{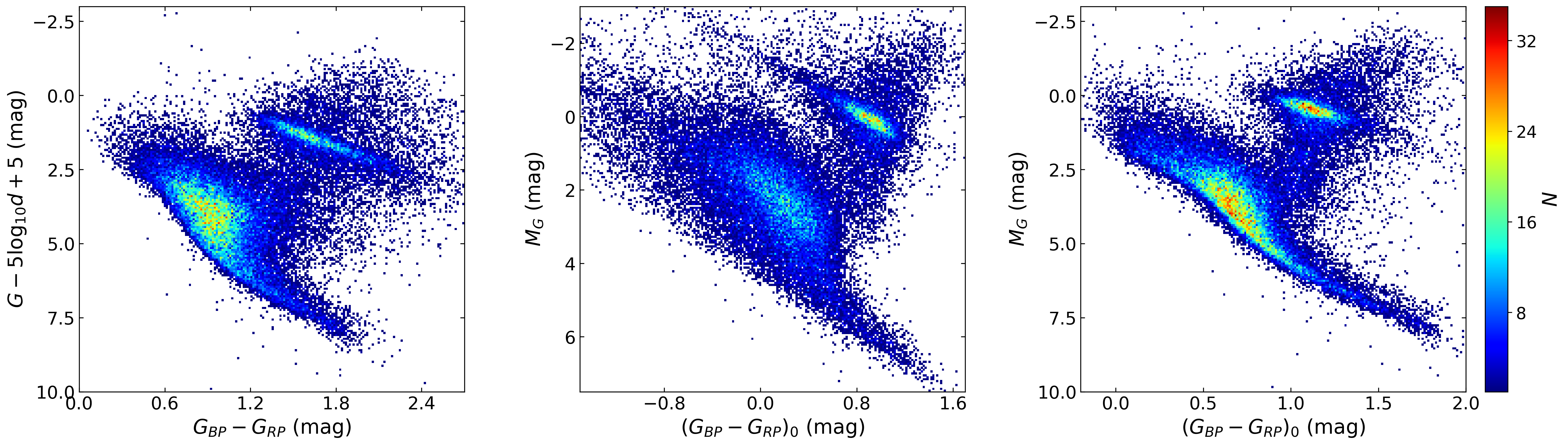}
\par\smallskip
\includegraphics[width=0.98\textwidth]{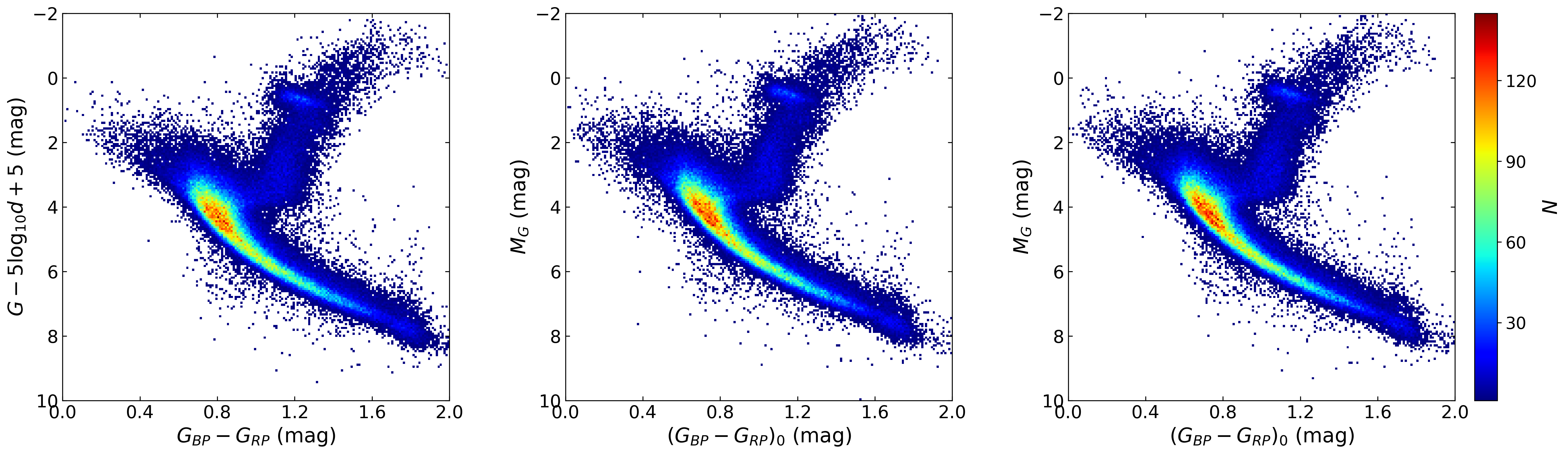}
\caption{
Color--magnitude diagrams before extinction correction (left), after correction with SFD (middle), and after correction with our three-dimensional extinction results (right). {\thickmuskip=3mu
The upper row shows the approximately $40~{\rm deg}^{2}$ field at $185.5^\circ < l < 191.5^\circ$ and $0.05^\circ < b < 6.73^\circ$. The lower row shows the approximately $200~{\rm deg}^{2}$ field at $112.5^\circ < l < 137.5^\circ$ and $-28.79^\circ < b < -20^\circ$.
} The three-dimensional reddening values are evaluated from the fitted distance--reddening profiles in HEALPix pixels with $N_{\rm side}=64$ at the distance of each star. The color scale indicates the number of stars per bin, $N$.}

\label{fig:color-absolute magnitude diagrams}
\end{figure}

To examine the application of our three-dimensional extinction results, we construct color--magnitude diagrams (CMDs) for two regions with different Galactic latitudes and extinction levels. 
The first covers approximately $40~{\rm deg}^{2}$ at $185.5^\circ < l < 191.5^\circ$ and $0.05^\circ < b < 6.73^\circ$. The second covers approximately $200~{\rm deg}^{2}$ at $112.5^\circ < l < 137.5^\circ$ and $-28.79^\circ < b < -20^\circ$.
For each star, the reddening is obtained by evaluating the fitted distance--reddening profile of its HEALPix pixel, with $N_{\rm side}=64$, at the adopted stellar distance, using the procedure described in Section~\ref{subsec:calculation of Line-of-Sight extinction profiles}. Figure~\ref{fig:color-absolute magnitude diagrams} compares the uncorrected CMDs with those corrected using the two-dimensional map of \citet{schlegel1998maps} (SFD) and our three-dimensional extinction results.

In the low-latitude field, our distance-dependent correction produces a better-defined main sequence (MS) and a more concentrated red clump (RC) than in the uncorrected CMD. The SFD correction instead produces a pronounced blueward displacement and a more dispersed MS. This behavior is consistent with excessive corrections for foreground stars when the full line-of-sight dust column represented by SFD is applied. In the higher-latitude field, the changes are more modest, and the CMDs corrected using SFD and our three-dimensional results are broadly similar. The corrected CMD of this approximately $200~{\rm deg}^{2}$ field shows a well-defined MS. These comparisons illustrate the benefit of accounting for stellar distance in the more heavily reddened field and provide an internal consistency check of our three-dimensional extinction results over different Galactic latitudes and angular areas.

\section{Conclusion}
\label{conclusion}

In this paper, we have compiled a combined, cross-calibrated stellar atmospheric parameter dataset comprising nearly ten million unique stars by integrating six major optical and near-infrared spectroscopic surveys (LAMOST, APOGEE-2, GALAH, Gaia-ESO, SEGUE, and DESI). Leveraging an enhanced multi-tier Star-Pair algorithm combined with an MCMC line-of-sight modeling framework, we have derived highly precise individual interstellar reddening values and mapped the 3D extinction distribution of the Galaxy. The primary findings and contributions of this work are summarized as follows:

\begin{enumerate}

    \item \textbf{A Combined Multi-Survey Parameter Catalog}: By adopting APOGEE-2 as the reference scale, we fitted MLPs and polynomials separately for giants and dwarfs to correct for systematic differences in $T_{\rm eff}$, $\log g$, and ${\rm [Fe/H]}$ across the complementary spectroscopic catalogs. Validation against the PASTEL high-resolution database and member stars in globular clusters shows good overall agreement. This delivers a value-added catalog of 9,862,056 unique stars with calibrated parameters, providing a useful foundation for both dust mapping and machine-learning parameter pipelines.

    \item \textbf{High-Precision Individual Extinction Estimates}: Using a 3D anisotropic parameter space indexed via \texttt{cKDTree}, we implemented a nested 4-tier adaptive decision algorithm (IDW and \texttt{Rbf}) to determine intrinsic colors for target stars. We derived $E(BP-RP)$ and $E(B-V)$ for 9,616,388 stars with a matching success rate of 97.43\%. Repeated observation tests on LAMOST multi-epoch stars demonstrate an internal reddening precision of $\sigma \approx 0.022~{\rm mag}$.

    \item \textbf{Robust 3D Extinction Mapping and Pipeline}: By modeling cumulative reddening as a monotonically increasing step function via MCMC and a robust Cauchy-inspired log-likelihood function, we established an efficient line-of-sight extinction fitting pipeline. Distance uncertainties are included in the effective reddening uncertainties used in the fitting. We constructed HEALPix-binned ($N_{\rm side}=64$) 3D extinction sky maps covering $33,894.87~{\rm deg}^{2}$, which trace prominent nearby molecular cloud complexes (e.g., Taurus, Perseus, Orion) as well as distant Galactic plane dust structures.

    \item \textbf{Comprehensive Benchmarking and Practical Applications}: Comparison with existing large-scale reddening maps (\citealt{wang2025all, zhang2025three, guo2021three, chen2019three}) reveals good overall agreement. Applying reddening corrections from the fitted line-of-sight profiles to regional color--magnitude diagrams produces a better-defined MS and a more concentrated RC in the low-latitude field. Comparison with SFD illustrates the benefit of distance-dependent corrections in this field, while the two corrections yield broadly similar CMDs in the approximately $200~{\rm deg}^{2}$ higher-latitude region.

\end{enumerate}

The derived individual interstellar reddening and extinction catalog from this work has been made publicly available at \url{https://nadc.china-vo.org/res/r101883/}. In addition, an open-source Python tool is provided alongside the catalog to facilitate user-defined spatial coordinate and distance queries for dynamic line-of-sight extinction extraction, serving as a valuable asset for future wide-field photometric calibration and Galactic evolution studies. Meanwhile, it also offers reliable data and tool support for the discovery and distance estimation of molecular clouds \citep[e.g.,][]{guo2022large, wang2025comprehensive}.

\normalem
\begin{acknowledgements}
This work is partially supported by the Physics Discipline Construction Project of Chuxiong Normal College (Grant No. 20260600016) and the Chuxiong Normal College Doctoral Research Initiation Fund Project under No. BSQD2586.

This work made use of the data from LAMOST (Large Sky Area Multi-Object Fiber Spectroscopic Telescope, also known as the Guoshoujing Telescope) (\url{https://cstr.cn/31118.02.LAMOST}). LAMOST is a Chinese national mega-science facility, operated by National Astronomical Observatories, Chinese Academy of Sciences.

Funding for the Sloan Digital Sky Survey IV has been provided by the Alfred P. Sloan Foundation, the U.S. Department of Energy Office of Science, and the Participating Institutions. SDSS acknowledges support and resources from the Center for High-Performance Computing at the University of Utah. The SDSS web site is \url{www.sdss4.org}.

This work has used the Fourth Data Release of the GALAH Survey \citep{buder2025galah}. The GALAH survey is based on data acquired through the Australian Astronomical Observatory, under the programmes: A/2013B/13 (The GALAH pilot survey); A/2014A/25, A/2015A/19, A2017A/18 (The GALAH survey, Phase 1); A2018A/18 (Open clusters with HERMES); A2019A/1 (Hierarchical star formation in Ori OB1); A2019A/15 (The GALAH survey, Phase 2); A/2015B/19, A/2016A/22, A/2016B/10, A/2017B/16, A/2018B/15 (The HERMES-TESS programme); and A/2015A/3, A/2015B/1, A/2015B/19, A/2016A/22, A/2016B/12, A/2017A/14 (The HER MES K2-follow-up programme). We acknowledge the traditional owners of the land on which the Anglo-Australian Telescope stands, the Gamilaraay people, and pay our respects to elders past and present. This paper includes data that has been provided by AAO Data Central \url{https://datacentral.org.au/}.

This work is based on data products from observations made with the ESO Telescope at the La Silla Paranal Observatory. These data products have been processed by the Cambridge Astronomy Survey Unit (CASU) at the Institute of Astronomy of Cambridge and by the FLAMES/UVES reduction team at INAF/Osservatorio Astrofisico di Arcetri. These data have been obtained from the GES Data Archive, prepared and hosted by the Wide Field Astronomy Unit, Institute of Astronomy, University of Edinburgh, which is funded by the UK Science and Technology Facilities Council. This work was partly supported by the European Union FP7 program through ERC grant No.320360 and by the Leverhulme Trust through grant RPG-2012541. We acknowledge the support of INAF and Ministero dell’Istruzione e Universitã e della Ricerca (MIUR) in the form of the grant “Premiale VLT 2012.” The results presented here benefit from the discussions held during the Gaia-ESO work shops and conferences supported by the European Science Foundation (ESF) through the GREAT Research Network Program. We also use data products created from observations collected at the European Organization for Astronomical Research in the Southern Hemisphere under the AMBRE Project, referring to the publications \citet{de2012ambre, de2013ambre} and \citet{worley2016ambre} and to the MATISSE parameterization algorithm \citet{recio2006automated}. The data are obtained from the ESO Science Archive Facility with doi:\url{10.18727/archive/25} and doi:\url{10.18727/archive/18}.

This research used data obtained with the Dark Energy Spectroscopic Instrument (DESI). DESI construction and operations is managed by the Lawrence Berkeley National Laboratory. This material is based upon work supported by the U.S. Department of Energy, Office of Science, Office of High-Energy Physics, under Contract No. DE–AC02–05CH11231, and by the National Energy Research Scientific Computing Center, a DOE Office of Science User Facility under the same contract. Additional support for DESI was provided by the U.S. National Science Foundation (NSF), Division of Astronomical Sciences under Contract No. AST-0950945 to the NSF’s National Optical-Infrared As
tronomy Research Laboratory; the Science and Technology Facilities Council of the United Kingdom; the Gordon and Betty Moore Foundation; the Heising-Simons Foundation; the French Alternative Energies and Atomic Energy Commission (CEA); the National Council of Humanities, Science and Technology of Mexico (CONAHCYT); the Ministry of Science and Innovation of Spain (MICINN), and by the DESI Member Institutions:\url{www.desi.lbl.gov/collaborating-institutions}. The DESI collaboration is honored to be permitted to conduct scientific research on I’oligam Du’ag (Kitt Peak), a mountain with particular significance to the Tohono O’odham Nation. Any opinions, findings, and conclusions or recommendations expressed in this material are those of the author(s) and do not necessarily reflect the views of the U.S. National Science Foundation, the U.S. Department of Energy, or any of the listed funding agencies.
\end{acknowledgements}
  
\bibliographystyle{raa}
\bibliography{bibtex}

\end{document}